\input{epsfig}
\documentstyle[prl,aps,twocolumn]{revtex}

\catcode`\@=11

\def\maketitle2{\par 
\begingroup
\let\cite\@bylinecite
\def\thefootnote{\fnsymbol{footnote}}%
\twocolumn[\@maketitle2\vskip2pc]%
\thispagestyle{plain}\@thanks
\endgroup
\def\thefootnote{\arabic{footnote}}%
\setcounter{footnote}{0}%
\let\maketitle2\relax \let\@maketitle2\relax
\let\@thanks\relax \let\@authoraddress\relax \let\@title\relax
\let\@date\relax \let\thanks\relax \let\@abstract\relax 
\let\@pacs\relax}

\def\abstract#1{\gdef\@abstract{{\par 
\bgroup
\ifdim\prevdepth=-1000pt \prevdepth0pt\fi
\hsize\columnwidth
\dimen0=-\prevdepth \advance\dimen0 by17.5pt \nointerlineskip
\small\vrule width 0pt height\dimen0 \relax}{~~}#1\egroup}}

\def\pacs#1{\gdef\@pacs{{\par 
\bgroup
\hsize\columnwidth \parindent0pt
\ifdim\prevdepth=-1000pt \prevdepth0pt\fi
\dimen0=-\prevdepth \advance\dimen0 by20pt\nointerlineskip
\egroup} PACS numbers:~#1}}

\def\@maketitle2{
\@preprint
\@title
\ifdim\prevdepth=-1000pt \prevdepth0pt\fi
\@authoraddress
\@date
\begin{list}{}{\leftmargin=0.10753\textwidth \rightmargin=\leftmargin
\itemsep=1pc\partopsep=-1pc}
\item\@abstract
\item\@pacs
\end{list}
}

\catcode`\@=12

\begin{document}
\draft
\preprint{LA-UR-97-4643}
\title{Decoherence, Chaos, and the Correspondence Principle}
\author{Salman Habib$^{1,\dagger}$, Kosuke Shizume$^{2,3,*}$, and
Wojciech Hubert Zurek$^{3,\#}$}  
\address{$^1$T-8, Theoretical Division, MS B285, Los Alamos National
Laboratory, Los Alamos, New Mexico 87545}
\address{$^2$University of Library and Information Science, 1-2
Kasuga, Tsukuba, Ibaraki 305, Japan} 
\address{$^3$T-6, Theoretical Division, MS B288, Los Alamos National
Laboratory, Los Alamos, New Mexico 87545} 
\date{\today}

\abstract 
{We present evidence that decoherence can produce a smooth
quantum-to-classical transition in nonlinear dynamical
systems. High-resolution tracking of quantum and classical evolutions
reveals differences in expectation values of corresponding
observables. Solutions of master equations demonstrate that
decoherence destroys quantum interference in Wigner distributions and
washes out fine structure in classical distributions bringing the two
closer together. Correspondence between quantum and classical
expectation values is also re-established.}


\maketitle2
\narrowtext

The status of quantum-classical correspondence for dynamical systems
is somewhat unclear and perhaps even controversial. In a nonlinear
system, a single classical trajectory initially centered on a quantum
wavepacket can quickly diverge from the motion of the centroid given
by the quantum expectation values of $x$ and $p$ (this defines the
Ehrenfest time, c.f. Ballentine {\em et al} in
Ref. \cite{ballref}). However, when the single trajectory is replaced
by a classical probability distribution to give the correspondence
principle a better chance, the situation is far less certain. Thus,
some authors have reported evidence for a breakdown of the
quantum-classical correspondence in chaotic systems \cite{corrbreak},
while others have argued that it can be preserved when stated in terms
of the expectation values of dynamical variables \cite{ballref}. In
yet another line of investigation it has been shown that, even in
chaotic systems, semiclassical methods are successful for times longer
than was previously believed possible \cite{hellref}.

In integrable systems, a rapid divergence between classical and
quantum evolutions can occur for initial conditions near an unstable
point: This can force the system to undergo a ``double-slit
experiment,'' yielding very different outcomes in the two cases. This
breakdown of correspondence is caused by the coherent interference of
fragments of the wavepacket and occurs on a short timescale, on the
order of the system dynamical time. However, for generic initial
conditions, quantum and classical expectation values are expected to
diverge on a timescale inversely proportional to some power of $\hbar$
\cite{BerZas,BerBal}. This is regarded as sufficiently slow to cause
no difficulties with the classical limit of quantum theory.

The first purpose of this Letter is to investigate the
quantum-classical correspondence at the level of expectation values
for chaotic dynamical systems, where a breakdown may be anticipated on
a much smaller {\em logarithmic} timescale $t_{\hbar}\sim\ln
(C/\hbar)$ \cite{BerZas,BerBal,ZurPaz}. We show below that such a loss
of correspondence {\em does} occur though its magnitude is not in
conflict with earlier results \cite{ballref}. While the timescale on
which the violation occurs varies with the particular initial
condition chosen, it is consistent with the logarithmic time
$t_{\hbar}$.

In order to affect a quantum-classical comparison we chose Gaussian
packets as initial states, with positive definite Wigner distribution
functions. The time evolution was then performed using either
classical or quantum dynamical equations. In the chaotic systems
investigated here, for generic initial conditions ({\em i.e.},
Gaussian packets randomly sampling the chaotic part of the phase
space), differences between quantum and classical expectation values
stay small for some time and then abruptly increase. After this
divergence time, the differences remain modest, typically $\sim
5-10\%$ (for $\hbar=0.1$). The Wigner function begins to differ
considerably from the classical phase space distribution at relatively
early times \cite{takaref}.

Our second purpose is to show that the discrepancy between quantum and
classical evolutions is drastically decreased by even a small coupling to
the environment, which in the quantum case leads to decoherence
\cite{deco}. There are two limiting situations in the study of
decoherence in dynamical systems. In the first case, special initial
conditions (such as Schr\"{o}dinger cat states) are used to study the
destruction of interference already present in the initial state but
with simple system dynamics, typically taken to be linear
\cite{linear}. However, since quantum interference is dynamically
generated only in nonlinear systems \cite{nonlin}, the competition
between generation and destruction of quantum coherence cannot be
investigated. The results reported in this Letter are from a study of
the second type, where the system dynamics is nonlinear, but the
choice of initial states is kept deliberately simple so as to focus
only on the role of dynamically induced interference (as distinct from
that present in the initial state). We show below that differences at
the level of expectation values are sharply reduced due to decoherence
and the effect on correspondence in phase space is even more
spectacular. In the parameter regime investigated here -- essentially
the border between quantum and classical -- some remnants of quantum
coherence may still survive. However, such small-scale coherence has
apparently little effect on the correspondence of the expectation
values.

A mechanism responsible for the quantum to classical transition should
explain not just how expectation values can converge to the same
answer, but also lead to compatible effective phase space
distributions. A common approach is an appeal to coarse-graining, a
formal procedure implemented typically by convolving the individual
distributions with a Gaussian and then comparing the two resulting
coarse-grained distributions \cite{takaref}. This approach has three
defects. First, as a formal mathematical procedure it can always be
inverted, and thus offers no physical insight. Second, this
coarse-graining does not alter the dynamics, and hence cannot improve
the convergence of expectation values. Third, for the classical
system, the notion of a trajectory is lost and along with it the
notion of Lyapunov exponent.

In contrast to the coarse-graining approach, decoherence 
provides a {\em dynamical} explanation \cite{deco} of the quantum to 
classical transition by taking into account interactions with an
(external or internal) environment of the system -- degrees of freedom
that effectively monitor and, therefore, select certain stable or
``pointer'' observables destined to become the classical variables
\cite{deco,hkmp}. The simplest models of this type lead to master
equations for the reduced density matrix for the system
\cite{masteq}. Diffusion terms in these equations automatically
coarse-grain the distributions, now a physical effect of the coupling
to the environment rather than a mathematical trick. The degree of
coarse-graining is determined by the interplay between the dynamics of
the system and the nature and strength of the coupling with the
environment. Moreover, the effectively classical master equations that
describe the post-decoherence dynamics admit a Langevin description of
trajectories allowing for the existence of a Lyapunov exponent. We
demonstrate below that decoherence dramatically improves the
correspondence of the expectation values, leads to the existence of a
single effective phase space distribution, and allows for a Lyapunov
exponent to exist at late times. All the deficiencies of the
coarse-graining approach are therefore overcome.

We restrict attention to bounded, one dimensional, driven
systems. Tools employed are very high-resolution simulations of the
time-dependent Schr\"{o}dinger, quantum and classical Liouville, and
master equations recently implemented on massively parallel computers
\cite{shrr}. The numerical results discussed below are for the driven
system considered in Ref. \cite{linbal}, with Hamiltonian,
\begin{equation}
H=p^2/2m + B x^4 - A x^2 + \Lambda x \cos(\omega t)~.
\label{lbham}
\end{equation} 
We used a parameter regime ($m=1$, $B=0.5$, $A=10$, $\Lambda=10$,
$\omega=6.07$) in which a substantial area of phase space is
predominantly stochastic, with the finite-time Lyapunov exponent
$\lambda \simeq 0.4 - 0.5$. Gaussian phase space distributions,
typically minimum uncertainty wavepackets, were employed as initial
conditions to sample the evolution in the stochastic region. Other
Hamiltonians studied include the driven Duffing system and a
two-dimensional, truncated Toda potential. These systems yielded
similar results; a detailed presentation will be given elsewhere
\cite{long}.

The specific model of decoherence used here is the weak coupling, high
temperature limit of quantum Brownian motion \cite{masteq}. In this
limit, dissipation can be ignored at early times, and only the
diffusive contributions in the master equations \cite{ZurPaz,deco}
need be kept. The diffusion constant was chosen to be small enough so
that, over the timescales of interest, the energy increase was
negligible, and changes in the classical phase space structure were
only perturbative. For the Hamiltonian (\ref{lbham}), the quantum
master equation in terms of the Wigner function is
\begin{equation} 
{\partial f_W \over \partial t}=-{p\over m}{\partial f_W \over \partial
x}+ {\partial V\over\partial x}{\partial f_W \over \partial p}+ L_q
f_W + D{\partial^2 f_W \over \partial p^2}~,
\label{qme} 
\end{equation}
where $\partial V/\partial x=4Bx^3-2Ax+\Lambda \cos(\omega t)$ and
$L_q$ is 
$$ 
L_q\equiv\sum_{n \ge 1}{\hbar^{2n}(-1)^{n} \over 2^{2n}(2n+1)!}
{\partial^{2n+1} V \over \partial x^{2n+1}}{\partial^{2n+1} 
\over \partial p^{2n+1}}=-\hbar^2Bx{\partial^3 \over \partial p^3}~.
$$
The classical ($L_q=0$) Fokker-Planck equation is
\begin{equation} 
{\partial f_c \over \partial t}=-{p\over m}{\partial f_c \over
\partial x}+ {\partial V\over\partial x}{\partial f_c \over \partial
p} + D{\partial^2 f_c \over \partial p^2}~.
\label{cme} 
\end{equation}
This dynamics is equivalent to the Langevin equation $m {\ddot
x}=-V'(x)+F(t)$ where $F(t)$ denotes a Gaussian, white noise. The
quantum Schr\"odinger and master equations, as well as the classical
Fokker-Planck equations were solved using a high-resolution spectral
algorithm \cite{shrr}. In the absence of diffusion, the classical
Liouville equation was solved as an N-body problem, the distribution
being sampled by at least $\sim 10^5$ particles. Numerical checks
included carrying out simulations at different spatial and temporal
resolutions and a direct verification for the moments obtained from
the codes by substituting them in the BBGKY-like moment evolution
hierarchy and verifying that this set of equations is satisfied to a
relatively high order. Lyapunov exponents were computed using the
techniques described in Refs. \cite{lyap}. It was verified that at the
noise levels used, the finite-time Lyapunov exponents from the
Langevin equation agreed with those computed from the Hamiltonian
dynamics.

The exponential instability characteristic of chaos forces the system
to rapidly explore large areas of phase space and to interfere on a
timescale set by when the wavefunction has spread over much of the
available space \cite{BerZas,BerBal,ZurPaz} and the Moyal corrections
arising from its nonlocality have become comparable to the classical
force \cite{ZurPaz}: 
\begin{equation} 
t_\hbar \sim \lambda^{-1}\ln(\chi\delta p/\hbar) ~,   \label{log}
\end{equation} 
where $\lambda$ is the Lyapunov exponent, $\delta p$ is the measure of
dispersion in the initial conditions, and
$\chi\simeq\sqrt{|\langle\partial_xV/\partial_{xxx} V\rangle|}$ is a
measure of the nonlinearity in the potential averaged over the
accessible space. We investigated the evolution of several expectation
values, such as $\langle x \rangle$, higher order moments such as
$\langle (x-\langle x\rangle)^n\rangle$ (with a maximum $n=4$), and
expectation values of variables that in principle include all
moments. Some of our results are shown in Fig. 1. In all cases, and
for all of the investigated initial conditions we found good agreement
between the quantum and classical results during the initial portion
of the evolution. (This initial period was {\em longer} than the
Ehrenfest time, consistent with the results of Ballentine {\em et al}
\cite{ballref}.) The onset of the discrepancy depended on the initial
condition, but in all cases was a factor of a few larger than the
dynamical time. The discrepancy saturates to typically no more than
$10\%$ of the expectation values thereafter. The value of $\hbar$ was
varied to test for logarithmic scaling: while the results are
consistent with (\ref{log}), the dynamic range of the simulations is
insufficient to make a more precise statement.

\vspace{.4cm}
\centerline{\epsfig{figure=figprl1.ps,height=6cm,width=8cm,angle=-90}}
\vspace{.35cm}
{FIG. 1. {\small{Classical and quantum expectation values $\langle
x\rangle$ as a function of time.  The initial condition is a minimum
uncertainty Gaussian wave packet with $\langle x\rangle=-3, \langle
p\rangle=8, \langle(x-\langle x\rangle)^2\rangle=0.0025, {\rm
and}~\langle(p-\langle p\rangle)^2\rangle=1$. $\hbar$ is set to
0.1. This yields $t_{\hbar}\sim 4$. The central vertical bar denotes
the average divergence time and the left and right bars, the minimum
and maximum respectively, for ten initial conditions that randomly
sampled the chaotic phase space.}}}\\

We have therefore good evidence that in isolated chaotic systems, the
quantum-classical correspondence defined at the level of expectation
values is lost relatively quickly due to dynamically generated quantum
interference. This is best appreciated by comparing classical phase
space densities with quantum Wigner distributions. As the wavepacket
spreads and folds, the Wigner function becomes dominated by small
scale interference which saturates on a scale set by the size of the
system in both momentum and position: $\delta p=\hbar/L$, $\delta
x=\hbar/P$. Eventually, the Wigner function is unable to track even
the ``backbone'' of the classical phase space distribution and becomes
a complicated looking interference pattern in which the classical
phase space structure can no longer be distinguished [Fig. 2(a)]. Fine
scale structure in the interference pattern ({\em i.e.}, oscillations
within an $\hbar$ box) is clearly apparent. On the timescales probed
in our simulations, the Wigner function has not reached the stage of
``structure saturation,'' {\em i.e.}, smoothness on a scale
$\sim\hbar$ in phase space \cite{BerBal}.

Our results are encapsulated in Figs. 2. As can be seen from comparing
the decohered Wigner function [Fig. 2(b)] with the classical
distribution given by the solution to Eq. (\ref{cme}) [Fig. 2(c)],
decoherence markedly improves the correspondence at the level of
distribution functions, radically changing the unitarily evolved
Wigner function of Fig. 2(a) by effectively smoothing it over scales
\cite{ZurPaz}; $\Delta p \simeq \sigma_c = \sqrt{2D/\lambda}$ in
momentum (which is translated by dynamics into a smoothing in
position). In our case $\sigma_c \simeq 0.3$ and $\chi\sim 0.6$, which
implies that we are on the border between quantum and classical
regimes (defined respectively by whether $\sigma_c \chi$ is large or
small compared to $\hbar$ \cite{ZurPaz}). Even though the
distributions in Figs. 2(b) and 2(c) are very close, the Wigner
function still contains traces of local quantum interference. However,
this makes only a minor difference to expectation values, and tends to
vanish as the evolution proceeds. We note that with our choice of
parameters, the diffusion term affects the evolution of classical and
quantum expectation values roughly to the same extent
(Fig. 1). Decoherence destroys the interference pattern in the Wigner
function, while at the same time, noise smooths out the fine structure
of the classical distribution in such a way that quantum and classical
distributions and expectation values both converge to each
other. Thus, one concludes that the decohered quantum evolution does
go over to the classical Fokker-Planck limit.

In summary, we have provided evidence that in a quantized classically
chaotic system, for fixed $\hbar$, classical and quantum expectation
values diverge from each other after a time $\sim t_{\hbar}$. In the
case studied here the discrepancy is $\leq 10\%$ of the typical
magnitude of the expectation values. Decoherence was shown to
substantially reduce this discrepancy as well as to bring the Wigner
and classical distributions very close to each other. In this combined
sense, decoherence restores the quantum-classical correspondence. Our
results complement previous studies which have focused more on the
phase space aspects of the correspondence and the destruction of
dynamical localization by noise and dissipation \cite{Ott}.

It is a pleasure to acknowledge discussions with J. R. Anglin,
C. Jarzynski, H. Mabuchi, S. Rugh, R. D. Ryne, B. Sundaram, and
G. M. Zaslavsky. KS acknowledges the Japanese Ministry of Culture and
Education for supporting his stay at Los Alamos National
Laboratory. Numerical simulations were performed on the CM-5 at the
ACL, LANL and the T3E at NERSC, LBNL.

\vspace{.4cm}
\centerline{\epsfig{figure=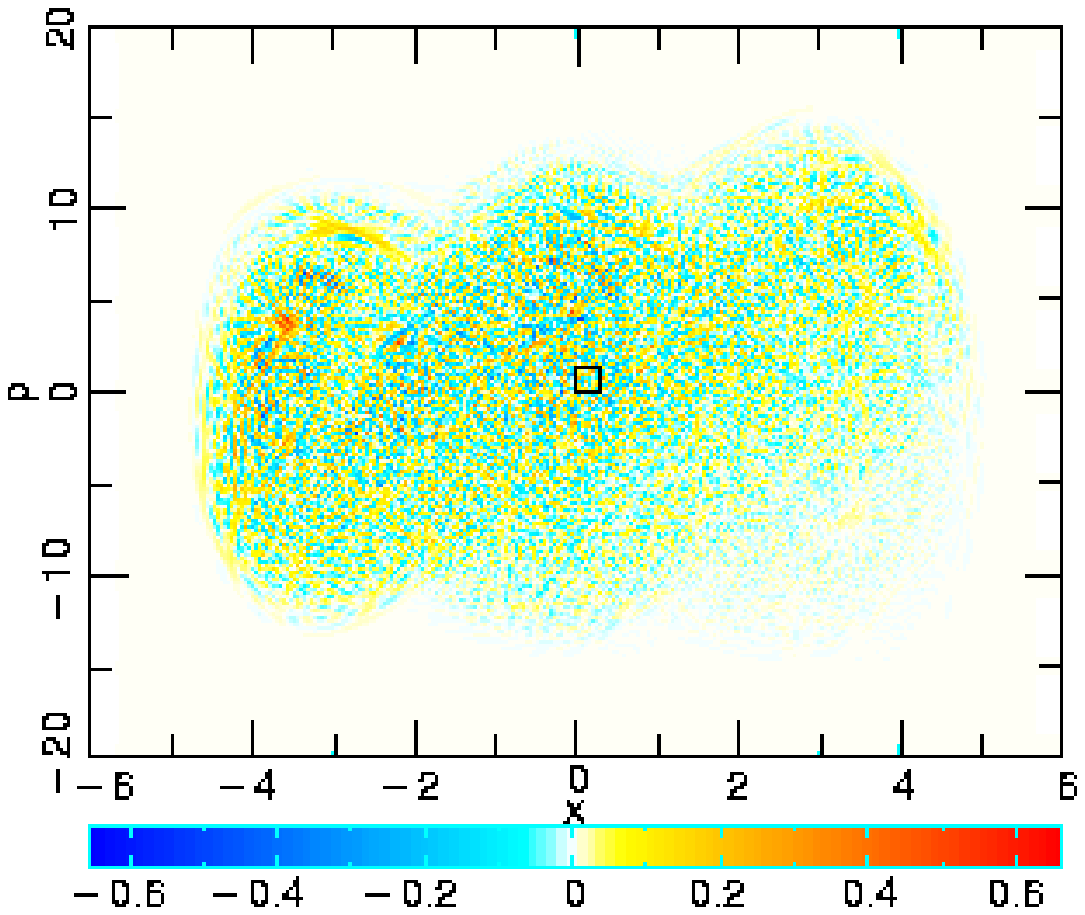,height=5cm,width=7cm,angle=0}}
\vspace{.35cm}
\centerline{\epsfig{figure=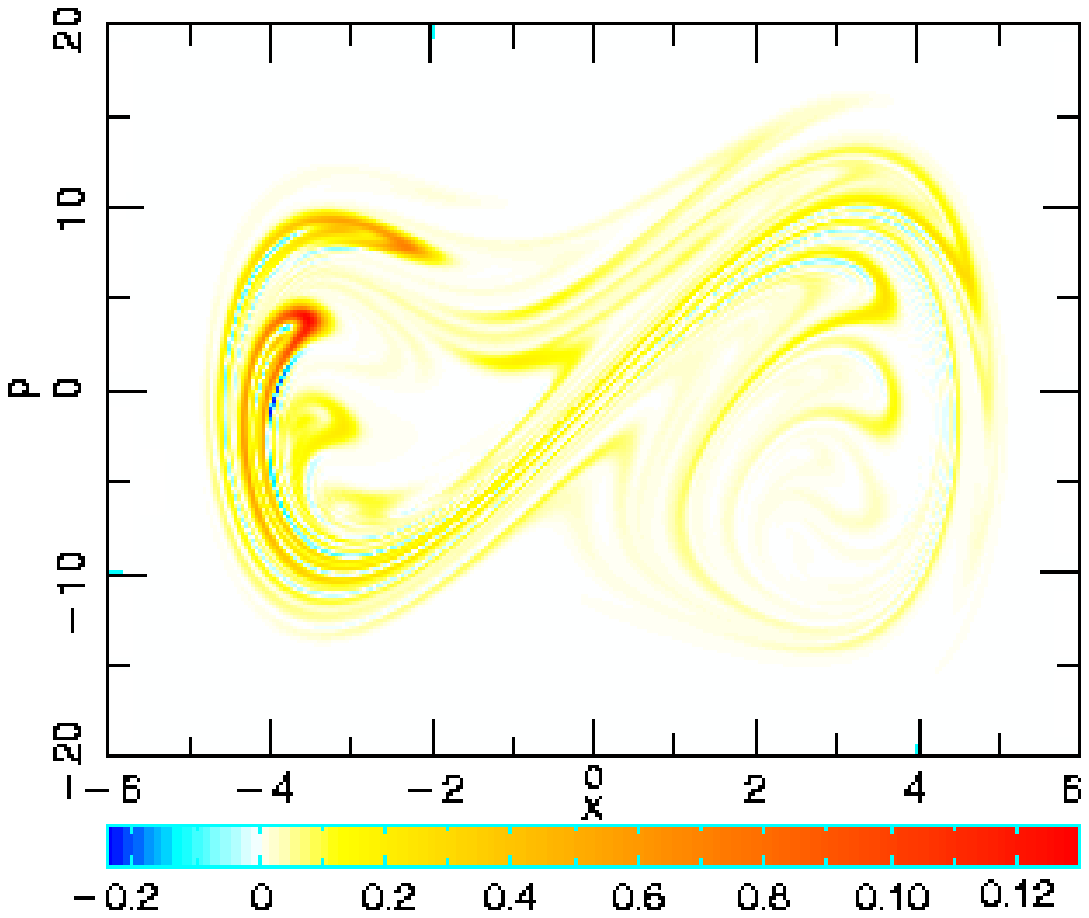,height=5cm,width=7cm,angle=0}}
\vspace{.35cm}
\centerline{\epsfig{figure=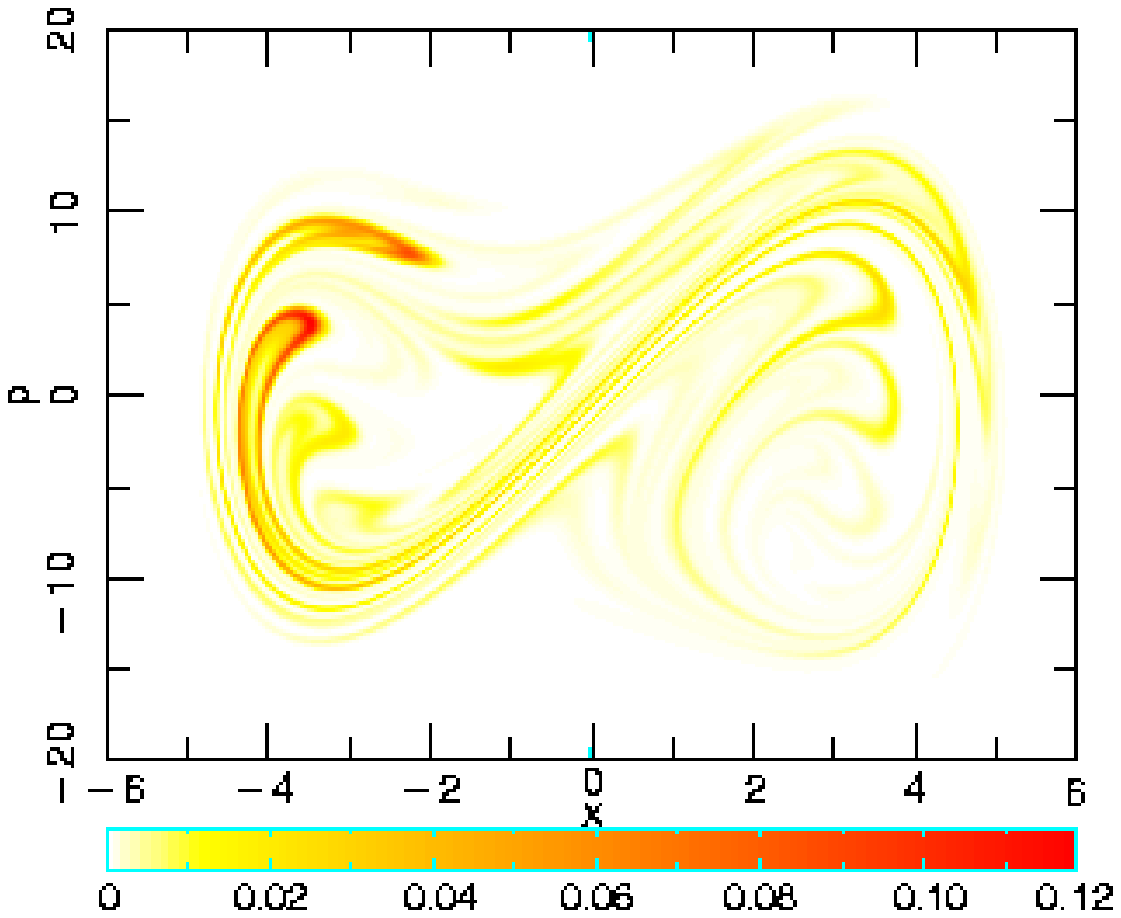,height=5cm,width=7cm,angle=0}}
\vspace{.35cm}
{FIG. 2. {\small{ (a) Wigner distribution function from a solution of
Eq. (\ref{qme}) at time $t=8T$, where $T$ is the period of the driving
force. The diffusion constant $D=0$. The box represents a phase space
area of $4\hbar$. (b) Wigner distribution function at time $t=8T$,
with diffusion constant $D=0.025$, illustrating the destruction of
large scale quantum coherence. (c) Classical distribution function
from a solution of Eq. (\ref{cme}) at time $t=8T$, with diffusion
constant $D=0.025$.}}}\\

\end{document}